\documentclass[preprint]{aastex}

\begin{document}
\title{An Upper Limit on the Mass of a Primordial Star\\ 
due to the Formation of an HII Region: \\
The Effect of Ionizing Radiation Force}
\author{Kazuyuki Omukai \altaffilmark{1,2} 
and Shu-ichiro Inutsuka \altaffilmark{3}}
\altaffiltext{1}{Division of Theoretical Astrophysics, 
National Astronomical Observatory, Mitaka, Tokyo 181-8588, Japan}
\altaffiltext{2}{Osservatorio Astrofisico di Arcetri, 
Largo E. Fermi 5, 50125 Firenze, Italy}
\altaffiltext{3}{Department of Physics, 
Kyoto University, Kyoto 606-8502, Japan}
\email{omukai@th.nao.ac.jp ; inutsuka@tap.scphys.kyoto-u.ac.jp}
\begin{abstract}
We analytically investigate the formation of an HII region 
in the accreting envelope of a newborn star.
Special care is taken to examine the role of ionizing radiation force.
This effect modifies velocity and density distributions and 
thereby affects the expansion of the HII region. 
As a result, the upper limit of the stellar mass imposed 
by the growth of an HII region around a forming star 
is increased by a larger factor than the previous estimate.
In particular, for a star forming out of metal-free gas, 
this mechanism does not impose a firm upper limit on its mass.
\end{abstract}


\section{Introduction}
Stars are supposed to be born in the dense regions of gas clouds.
The theory of the gravitational fragmentation of the gas clouds
describes how these dense regions are formed inside
the parental gas clouds.
The typical mass scale of the dense region can be predicted by
the Jeans mass of the cloud at the time of the fragmentation.
However, a perturbation with a mass scale larger than
this Jeans mass is always unstable.
Therefore a dense region that is more massive than the typical
Jeans mass can form depending on the initial fluctuation spectrum
in the cloud.
In this sense, the upper limit for the mass of stars cannot be
obtained from the analysis of the gravitational fragmentation.

Once, the origin of the observed upper limit of stellar mass 
(around $100M_{\sun}$) was attributed to the instability 
of massive stars due to the $\varepsilon$-mechanism 
(e.g., Schwarzshild \& H\"{a}rm 1959). 
Since later studies revealed that this mechanism leads to 
moderate mass-loss rather than disruption of the stars 
(e.g., Appenzeller 1970), formation process, 
instead of the stability of massive stars, 
has been believed to limit the upper bound for the stellar mass
(e.g., Nakano, Hasegawa, \& Norman 1995).

The star formation process is an accretion of 
ambient matter onto a protostar (stellar core) forming 
inside a protostellar cloud.
In this scenario, the final mass of a star depends on
how much mass the protostar can acquire.
Although it is not yet clear what mechanism stops the accretion,
increasingly strong feedback from a massive forming star 
is likely to halt the accretion.

Larson \& Starrfield (1971) were first to propose 
that a firm upper limit on the stellar mass is provided 
by the formation of an HII region around a
forming star, as well as by the radiation force acting on dust grains
in the accreting envelope.
The formation of an HII region prevents further accretion 
in the following way:
when the HII region reaches outer layers of a protostellar cloud, 
the temperature and then the pressure support surge by some large 
factor, so that the further infall of material is immediately halted.
Larson \& Starrfield (1971) concluded that both of those mechanisms 
set an upper limit of about 50 M$_{\sun}$ 
on the masses of Pop I stars 
forming out of the present interstellar medium.
The radiation force onto dust grains was found by subsequent studies 
 to set more stringent mass upper limit
(Appenzeller \& Tscharnuter 1974; Kahn 1974; 
Yorke \& Kr\"{u}gel 1977; Wolfire \& Cassinelli 1987).
Among them, Wolfire \& Cassinelli (1987) suggested that even stars as
massive as about 10 M$_{\sun}$ cannot form by accretion 
owing to this mechanism unless dust is significantly depleted.

However, for so-called Pop III stars forming out of metal-free gas, 
this mechanism of the radiation force does not work 
because of the absence of dust grains.
The $\varepsilon$-mechanism drives even milder mass-loss 
for those stars than Pop I massive stars (Baraffe, Heger, \& Woosley 2001).
According to recent simulations of the fragmentation of primordial 
clouds, the mass scale of fragments (protostellar clouds) is as large as 
$10^{3} M_{\sun}$ (Bromm, Coppi, \& Larson 1999;
Abel, Bryan, \& Norman 2000; Tsuribe 2001). 
Therefore, the growth of the HII region plays a crucial role 
in limiting the maximum mass of a Pop III star, 
if the upper limit by this mechanism is less than the mass scale of fragments.
This motivates us to study here the upper mass limit of Pop III stars due 
to the formation of an HII region.

In discussing the formation of an HII region, 
often the free-fall assumption has been imposed on the flow 
in the accreting envelope (e.g., Yorke 1986).
The role of the momentum transfer to the gas due to ionizing radiation 
has been neglected.
In the context of galaxy formation,  
however, its significance in dynamical and thermal evolution 
of the intergalactic medium has been pointed out by Haehnelt (1995).
He showed, for the collapse of subgalactic clouds of $\la 10^{10} M_{\sun}$,
even a radiation-pressure-driven bounce is possible.
Taking this effect into account, in this paper, we study 
the formation of an HII region in the accreting envelope of 
a forming massive star.
Although this effect is negligible as long as the radius of 
the HII region is small, 
it will turn out that inclusion of the ionizing radiation force 
alters the later evolution of the HII region.
In the course of the expansion of the HII region, the flow becomes 
slower than the free-fall rate owing to the ionizing radiation force. 
The growth of the HII region is strongly suppressed 
by this effect.
Consequently, the mass upper limit imposed by the expansion of 
the HII region is increased by a large factor.
In particular, the formation of an HII region does not set 
any firm mass upper limit for Pop III stars.

In Section 2, we briefly summarize the relevant aspects of the formation 
of an HII region in a free-falling envelope around a newborn star.  
The effect of ionization radiation force is included in Section 3.
Finally, we provide a summary and discussion in Section 4.

\section{Formation of an HII Region in a Free-Falling Envelope}
In this section, we review the expansion law of an HII region 
around a newly formed star (e.g., Yorke 1986) 
and discuss the resultant upper mass limit of a forming star
assuming that the accreting flow is free-falling.

Suppose an accreting flow onto a star of mass $M_{\ast}$ is taking place
at an accretion rate $\dot{M}$.
The mass accretion rate $\dot{M}$ is determined by the temperature
of the protostellar cloud 
and does not change very much except during the very late stage of 
the accretion phase where most of the mass of the envelope 
has accreted onto the central star.
The relation between $\dot{M}$ and the sound speed $c_{\rm s}$ in the 
protostellar cloud is given by 
\begin{equation}
\dot{M} = C_{\rm acc} \frac{c_{\rm s}^{3}}{G},  
\end{equation}
where $C_{\rm acc}$ is a non-dimensional number 
(see, e.g., Whitworth \& Summers 1987). 
When the numerical value of $\dot{M}$ is needed, 
we use $10^{-3}M_{\sun}{\rm yr}^{-1}$ 
corresponding to $C_{\rm acc}=1$ and 400K 
which is a typical temperature of metal-free protostellar clouds
(e.g., Yoneyama 1972; Carlberg 1981; Palla, Salpeter, \& Stahler 1983).

Here and throughout this paper, we assume that the accreting flow is 
spherically symmetric and steady.
Then, the continuity equation yields
\begin{equation}
\label{eq:steady}
\dot{M}=-4 \pi r^{2} \rho u,
\end{equation}   
$\rho$ and $u$ are the density and velocity in the accreting flow 
at radius $r$.
In addition, we assume in this section 
that the accreting envelope is free-falling: 
\begin{equation}
\label{eq:freefall}
u=-\sqrt{\frac{2GM_{\ast}}{r}}.
\end{equation}
Using equations (\ref{eq:steady}) and (\ref{eq:freefall}), 
we obtain the density distribution 
\begin{equation}
\label{eq:denslaw}
\rho=\frac{\dot{M}}{4 \pi \sqrt{2G M_{\ast}}} r^{-3/2}.
\end{equation}

In the HII region, we assume that the ionization and recombination are
in equilibrium and that all the re-emitted ionization photons are 
consumed locally (a so-called on-the-spot approximation).
Suppose $Q$ ionizing photons are emitted per unit time from a sphere 
of radius $R_{\rm in}$ surrounding the star.
We regard the stellar surface 
at $R_{\rm in}$ 
as the base of the HII region 
for the moment (see discussion in \S 4).
For the sake of simplicity, we assume that the gas is purelycomposed of  
hydrogen and is fully ionized inside the HII region
(i.e., the number density of H$^{+}$ $n({\rm H^{+}})=n(e)=\rho/m_{\rm p}$), 
hereafter.
We obtain the radius $R_{\rm II}$ of the HII region
from the ionization equilibrium:
\begin{equation}
\label{eq:ioneq}
Q=\int_{R_{\rm in}}^{R_{\rm II}} \alpha  n({\rm H^{+}})n(e) dV 
+ \frac{\dot{M}}{m_{\rm p}},
\end{equation}
where the recombination coefficient $\alpha=\alpha_{\rm B}$ 
from the on-the-spot approximation.
The first term on the right hand side means the recombination rate 
in the HII region.
The second term means the neutral flux entering into the HII region.  
Note that no HII region forms if the second term on the right-hand side 
is larger than $Q$.
In this paper, we neglect the $\dot{M}/m_{\rm p}$ term, 
assuming $Q \gg \dot{M}/m_{\rm p}=3.8 \times 10^{46} 
\dot{m}_{-3} ({\rm s}^{-1})$, 
where $\dot{m}_{-3} \equiv \dot{M}/10^{-3}M_{\sun}{\rm yr}^{-1}$.
Using the relation (\ref{eq:denslaw}), we obtain
\begin{equation}
Q= \frac{\alpha {\dot{M}}^{2} }{8 \pi G {m_{\rm p}}^{2} M_{\ast}} 
{\rm ln} (R_{\rm II}/R_{\rm in}). 
\end{equation}
Then, 
\begin{equation}
\label{eq:explaw}
R_{\rm II}=R_{\rm in} {\rm exp}(Q/Q_{\rm crit}^{\rm (FF)}),
\end{equation}
where
\begin{eqnarray}
Q_{\rm crit}^{\rm (FF)}
&=& \frac{\alpha {\dot{M}}^{2}}{8 \pi G {m_{\rm p}}^{2} M_{\ast}} \\
&=& 1 \times 10^{51} (M_{\ast}/100M_{\sun})^{-1} {\dot{m}_{-3}}^{2} 
({\rm s}^{-1}).
\label{eq:qcrit}
\end{eqnarray}
This shows that if $Q$ exceeds $Q_{\rm crit}^{\rm (FF)}$, 
the HII region expands exponentially.
Soon after, the HII region reaches an outer layer 
whose gravitational binding energy 
is lower than the thermal energy of ionized matter.
This results in the immediate expulsion of the outer layer 
and the halting of the accretion.

When does the ionizing photon emissivity $Q$ of the star 
reach the critical value $Q_{\rm crit}^{\rm (FF)}$?
To be concrete, we consider here the case of Pop III stars.
The ionizing photon emissivity
\begin{equation}
\label{eq:qedd}
Q \simeq 1.6 \times 10^{50} (M_{\ast}/100M_{\sun}) ({\rm s^{-1}})
\end{equation}   
for very massive ($M_{\ast} \ga 300 M_{\sun}$) Pop III stars in 
the zero-age main sequence (ZAMS; Bromm, Kudritzki, \& Loeb 2001).
Comparing (\ref{eq:qedd}) with (\ref{eq:qcrit}), 
we obtain the critical stellar mass above where $Q$ exceeds 
$Q_{\rm crit}^{\rm (FF)}$:
\begin{equation}
M_{\ast} \simeq 3 \times 10^{2} \dot{m}_{-3} M_{\sun}.
\end{equation}
The formation of a more massive star is 
increasingly difficult because of the growth of the HII region.

Recall here that the exponential expansion law of the HII region 
(\ref{eq:explaw}), which plays a key role in determining the upper limit 
of stellar mass, is a direct consequence of the density distribution 
(\ref{eq:denslaw}) and then the free-fall assumption (\ref{eq:freefall}).
However, the gas is subject to momentum transfer 
due to ionizing radiation and becomes slower than the free-fall rate
within the HII region.
This modifies the density law (\ref{eq:denslaw}) and the 
expansion law of the HII region (\ref{eq:explaw}).
Taking this effect into account, we study, in the next section, 
the growth of an HII region in the accreting envelope of 
a newborn massive star.

\section{Effect of Radiation Force on Expansion of an HII Region}
Hydrogen atoms in the HII region are ionized both by ionizing photons 
emitted directly by the star,
and by re-emitted ionizing photons due to the recombination 
to the ground level.
While the former is radially outward-directed, 
the re-emitted photons become almost isotropic and 
contribute little to the radiation force in the radial direction.
For the calculation of radiation force,
we must subtract 
the contribution from the ionization by locally emitted photons  
from the total ionization rate $\beta n({\rm H})$.   
The rate of the former equals 
the recombination rate to the ground level 
$\alpha_{1} n(\rm{H^{+}}) n(e)$ (per unit volume) 
according to the on-the-spot approximation,
while the latter equals the total recombination rate
$\alpha_{\rm A} n({\rm H^{+}})n(e)$ by ionization equilibrium.
Hence, the radiation force by ionization radiation per unit mass
is given by
\begin{equation}
f_{\rm ion} = \frac{h \nu_{\rm ion}}{c} \alpha n({\rm H^{+}})n(e)/ \rho ,
\end{equation}
where, $\alpha=\alpha_{\rm B}$ again.
Since we are considering stellar UV sources of thermal spectra,
the mean energy of ionizing photons is 
$h \nu_{\rm ion} \simeq h \nu_{0} =13.6$eV.  
For later convenience, we define the Haehnelt's radius
\begin{equation}
R_{\rm H}=\frac{\alpha h \nu_{\rm ion}}{4 \pi c G m_{\rm p}^{2}} 
\sim 26 {\rm pc},
\end{equation}
first introduced by Haehnelt (1995),
which is a fundamental length scale for the balance between 
ionizing radiation force and gravity.\footnote{
Comparing radiation force 
$f_{\rm ion}=4 \pi G R_{\rm H} \rho$ from a point source with gravity 
$f_{\rm grav}=G(4 \pi \rho r^{3}/3)/r^{2}$ exerted by homogeneously distributed
matter with density $\rho$ around the source, 
Haehnelt (1995) found the radiation force dominates the gravity 
at $r<3R_{\rm H}$.}
Finally, the expression of the ionizing radiation force becomes
\begin{equation}	
f_{\rm ion} = 4 \pi G R_{\rm H} \rho.
\end{equation}
Note that we have neglected other photon extinction processes  
such as 
electron scattering, photodissociation of H$^{-}$, etc. 
This assumption is valid as long as the HII region outside $R_{\rm in }$ 
is optically thin to those extinction processes.

We consider here an HII region with the steady radius $R_{\rm II}$
determined by the ionization equilibrium (\ref{eq:ioneq}) as 
in the previous section.
The I-front moves outward slowly as the stellar mass, and then  
the ionization photon emissivity from the star, increase. 
If the flow velocity across the ionization front (I-front) is more 
than twice the sound speed of ionized matter, 
the I-front becomes a (weak) R-type (Kahn 1954).
Since we consider the steady radius of the HII region, the velocity across
the I-front equals that of the flow at the edge of the HII region.
As long as the HII region is not large, 
the flow velocity is supersonic and fast enough to keep the I-front 
an R-type.
In fact, as we will show later, the halting of accretion occurs while
the I-front is still R-type. 
Hence, we treat here only the HII region bounded by R-front and do not 
consider later evolution.

Including the ionization radiation force, the momentum equation 
for the steady flow yields
\begin{equation}
\label{eq:momentum}
u \frac{du}{dr}=-\frac{GM_{\ast}}{r^{2}}+ 4 \pi G R_{\rm H} \rho 
\end{equation}
in the HII region ($R_{\rm in} \leq r \leq R_{\rm II}$).
The thermal pressure term can be neglected since we consider 
only supersonic flow.

Using the density-velocity relation in the steady accretion 
\begin{equation}
                  \rho = \frac{\dot{M}}{4 \pi r^2} (-u)^{-1},
\end{equation}   
and introducing the kinetic energy density of the 
flow $K \equiv u^{2}/2$, we obtain from equation (\ref{eq:momentum})
\begin{equation}
\label{eq:flow}
\frac{dK}{dr}=-\frac{GM_{\ast}}{r^{2}} 
                \left( 1-\sqrt{ \frac{K_{\rm crit}}{K}} \right). 
\end{equation}
Here we define the critical kinetic energy $K_{\rm crit}$ and the critical
flow velocity $u_{\rm crit}$:
\begin{equation}
\label{eq:Kcrit}
K_{\rm crit}=\frac{1}{2} u_{\rm crit}^{2} 
=\frac{1}{2}(R_{\rm H}/t_{\rm acc})^{2},
\end{equation}
and the accretion timescale
\begin{equation}
\label{eq:tacc}
                       t_{\rm acc}=M_{\ast}/\dot{M}.
                       \label{eq:accretiontime}
\end{equation}

Given the value of $K(=K_{\rm II})$ at the outer boundary $R_{\rm II}$ 
of the HII region, 
we can solve the momentum equation (\ref{eq:flow}) inward.
Depending on whether the initial value $K_{\rm II}$ is larger than,
equal to, or smaller than the critical value $K_{\rm crit}$, 
solutions of the equation (\ref{eq:flow}) are classified into three types.
In Figure 1, the solutions for $K_{\rm II}/K_{\rm crit}=
1.1, 1.01, 1, 0.99$, and  0.9 are shown. 
If $K_{\rm II}$ is larger than the critical value $K_{\rm crit}$, 
the right hand side of (\ref{eq:flow}) is always negative.
Then $K$ becomes larger and larger as the gas flows inward: 
The flow is accelerated.
We call these types of solutions accelerating solutions.
If $K_{\rm II}=K_{\rm crit}$, 
$K$ remains always at this value
(the uniform velocity solution), 
\begin{equation}
u(r) = u_{\rm crit} = - \frac{R_{\rm H}}{t_{\rm acc}}
\label{eq:uniform-v}
\end{equation}   
In other words, the ionizing radiation force exactly balances
the gravity everywhere in the HII region.
From equations (\ref{eq:steady}), (\ref{eq:Kcrit}) and (\ref{eq:tacc}), 
the density distribution of the uniform velocity solution is given by
\begin{equation}
\label{eq:rhoc}
\rho=\frac{M_{\ast}}{4 \pi R_{\rm H}} r^{-2}. 
\end{equation}
Similarly, if $K_{\rm II}<K_{\rm crit}$, the flow continues
 to decelerate (decelerating solutions).
As can be seen in Figure 1, the deceleration is so rapid 
that the accretion is halted by the radiation force 
without reaching the star shortly after the 
critical flow is attained.

As the outer boundary condition, 
we assume that the flow is free-fall outside the ionizing region.
We further assume that the flow velocity is continuous at the I-front,
since velocity jump at the R-front is small. 
Then, 
\begin{equation}
\label{eq:bc}
K_{\rm II}=\frac{GM_{\ast}}{R_{\rm II}} , 
\end{equation}
the kinetic energy $K_{\rm II}$ at the outer boundary of the HII region 
is related to the radius $R_{\rm II}$ of the HII region. 

The flow equation (\ref{eq:flow}) can be solved analytically.
The solution satisfying the boundary condition (\ref{eq:bc}) is
\begin{equation}
\frac{GM_{\ast}}{r}=K
+2 \sqrt{K_{\rm crit}} (\sqrt{K}-\sqrt{K_{\rm II}})
+2 K_{\rm crit} 
{\rm ln}(\frac{\sqrt{K}-\sqrt{K_{\rm crit}}}
{\sqrt{K_{\rm II}}-\sqrt{K_{\rm crit}}})
~~({\rm if}~K_{\rm II} \neq K_{\rm crit}),
\label{eq:exact}
\end{equation}
and
\begin{equation}
K=K_{\rm crit}~~({\rm if}~K_{\rm II}=K_{\rm crit}).
\end{equation}
In the limit of small radiation force (i.e., $K_{\rm crit} \rightarrow 0$),
only the first term on the right hand side of equation (\ref{eq:exact})
remains and the flow becomes free-fall. 
The second and third terms are corrections due to the radiation force.

From the above characteristics of the flow equation (\ref{eq:flow}), 
the growth of the HII region can be described as follows:
As long as the the forming star is not massive enough to emit a large quantity 
of ionization photons, the HII region around it is compact.
Since the radius $R_{\rm II}$ is small, $K_{\rm II} > K_{\rm crit}$
in this stage.
The flow inside the HII region is hardly affected by the ionizing radiation 
force.
As the star grows in mass and begins to emit a large quantity 
of ionizing photons, the radius of the HII region becomes larger and larger.
When the radius of the HII region reaches the critical value 
\begin{equation}
\label{eq:rcrit}
R_{\rm II crit}=\frac{2GM_{\ast}}{(R_{\rm H}/t_{\rm acc})^{2}}
=0.59 \times 10^{3} R_{\sun} (M_{\ast}/100M_{\sun})^{3} {\dot{m}_{-3}}^{-1},
\end{equation}
a uniform velocity flow is attained inside the HII region.
Further growth of the HII region would result in non-steady accretion
 and the accretion would eventually halt.
We cannot discuss the later evolution of the accreting flow 
within the framework of steady accretion. 
According to hydrodynamical calculations for the accretion being halted 
by the radiation force onto dust,  
the accretion rate begins to decrease with some oscillations 
when the luminosity reaches the critical value (Yorke \& Kr\"{u}gel 1977).
We speculate that a similar phenomenon would occur also in our case.

We justify here the assumption that the I-front remains in R-type 
until it reaches the critical radius $R_{\rm II crit}$.
When the radius of the HII region reaches $R_{\rm II crit}$, 
the flow velocity in the HII region is 
$u_{\rm crit}$ (see eq. \ref{eq:uniform-v}).
The I-front is R-type if 
\begin{equation}
|u_{\rm crit}|> 2  c_{\rm s II},
\end{equation} 
where $c_{\rm s II}$ is the sound speed of the ionized matter.
Then, our assumption of R-front is justified as long as 
\begin{equation}
\label{eq:Rcond}
                 t_{\rm acc} < \frac{R_{\rm H}}{2 c_{\rm s II}} 
                               \simeq 2 \times 10^{6} {\rm yr},
\end{equation}
where for $c_{\rm s II}$ we used the value at 10$^{4}$K.
The right hand side of equation (\ref{eq:Rcond}) is as large 
as the lifetime of a massive star.
Recall that the accretion time $t_{\rm acc}$ means 
the time needed to form a star of mass $M_{\ast}$ 
at an accretion rate $\dot{M}$.
Since the halting must occur within the stellar lifetime, 
the condition (\ref{eq:Rcond}) is always met. 
Therefore, in our case, we can conclude that the I-front remains R-type
until the halting of the accretion.
If the mass accretion rate decreases significantly, 
the formally defined accretion timescale $t_{\rm acc}$ 
(eq.[\ref{eq:accretiontime}]) can be larger than the stellar lifetime, 
and hence, the condition (\ref{eq:Rcond}) 
would be violated within the stellar lifetime.  
However this decrease of the mass accretion rate is expected 
only in the very late stage of the mass accretion phase where 
the mass of the envelope is significantly depleted.  
Therefore the mass of the star eventually becomes on the order of the mass 
of the parental cloud.  
That is, the halting mechanism due to the ionizing photons considered here 
is not important in this case.

Next, we estimate the ionizing photon emissivity needed 
for attainment of the uniform velocity solution.
Substituting the density distribution at the critical flow (\ref{eq:rhoc})
into the equation of ionization equilibrium (\ref{eq:ioneq}), we obtain 
the critical ionization photon emissivity
\begin{eqnarray}
\label{eq:qc_0}
Q_{\rm crit} &=& \frac{\alpha \dot{M}^{2}}{8 \pi m_{\rm p}^{2} K_{\rm crit}}
(\frac{1}{R_{\rm in}}-\frac{1}{R_{\rm II crit}}) \\
&=& (\frac{h \nu_{\rm ion}}{c})^{-1} \frac{G M_{\ast}^{2}}{R_{\rm H}}
(\frac{1}{R_{\rm in}}-\frac{1}{R_{\rm II crit}}),
\end{eqnarray}
above in which the flow is decelerated in the HII region.
We have neglected the accretion term $\dot{M}/m_{\rm p}$ as in \S 2.

In equation (\ref{eq:qc_0}), if the right hand side is 
dominated by the first term (i.e., $R_{\rm in} \ll R_{\rm II crit}$), 
\begin{equation}
Q_{\rm crit}=(\frac{h \nu_{\rm ion}}{c})^{-1} 
\frac{G M_{\ast}^{2}}{R_{\rm in} R_{\rm H}}
=0.64 \times 10^{53} (R_{\rm in}/10R_{\sun})^{-1} 
(M_{\ast}/100M_{\odot})^{2}.
\end{equation}
As a typical value of $R_{\rm in}$, we take 10$R_{\sun}$, which is 
approximately the radius of ZAMS Pop III stars of 500$M_{\sun}$
(El Eid, Fricke, \& Ober 1983).

Although we describe both $Q_{\rm crit}$ (eq. [\ref{eq:qc_0}]) and 
$Q_{\rm crit}^{\rm (FF)}$ in \S 2 (eq. [\ref{eq:qcrit}]) 
as the critical ionizing photon emissivities, 
mechanisms of halting the accretion are different.
In \S 2, the accretion is halted by the gas pressure in the HII region,
whereas here what halts the accretion is the radiation pressure of 
ionizing photons.
The inclusion of ionization radiation force
results in enhancement of $Q_{\rm crit}$
because of an increase in the density (and hence, the recombination
rate) due to the reduced flow velocity.

So far, we have only considered ionizing radiation as a source 
of radiation force.
However, for massive stars the luminosity approaches the Eddington limit,
and then the radiation force due to electron scattering becomes important.
We can include this effect in our scheme without difficulty.
If we use $G_{\rm eff}=G(1-\Gamma)$ and $R_{\rm H,eff}=R_{\rm H}/(1-\Gamma)$ 
(where $\Gamma=L/L_{\rm Edd}$) instead of $G$ and $R_{\rm H}$,  
the flow equation including the radiation force due to electron scattering 
retains the same form as equations (\ref{eq:momentum}) and (\ref{eq:flow}).
Note that $G$ in (\ref{eq:bc}) is not replaced 
since the gravity outside the HII region is not altered.
After the same procedure as that leading to (\ref{eq:qc_0}), 
the critical emissivity of ionization photons becomes 
\begin{equation}
\label{eq:qc_es}
Q_{\rm crit}=(\frac{h \nu_{\rm ion}}{c})^{-1} \frac{G M_{\ast}^{2}}{R_{\rm H}}
[\frac{(1-\Gamma)^{2}}{R_{\rm in}}-\frac{1}{R_{\rm II crit}}].
\end{equation}

The critical ionizing photon emissivities $Q_{\rm crit}$ given by
equations 
(\ref{eq:qc_0}) and (\ref{eq:qc_es}), namely the critical values with and
without the effect of the electron scattering, are both depicted in Figure 2
as functions of the central stellar mass.
In this figure, we have used 
$\dot{M}=10^{-3} M_{\sun}{\rm yr}^{-1}$ and $R_{\rm in}=10R_{\sun}$ as
before.
Also shown are the ionizing photon emissivity $Q$ of 
ZAMS Pop III stars and the critical value 
$Q_{\rm crit}^{({\rm FF})}$ obtained by assuming the free-fall 
(eq. [\ref{eq:qcrit}]). 
While $Q$ exceeds $Q_{\rm crit}^{\rm (FF)}$ at about 300$M_{\sun}$ 
as already mentioned in \S 2, $Q_{\rm crit}$ is about two orders 
of magnitude larger than the actual value $Q$ even at $10^{3}M_{\sun}$
for our assumed value of $R_{\rm in}=10R_{\sun}$,
even though the electron scattering reduces $Q_{\rm crit}$ 
considerably. 
Recalling that $Q_{\rm crit} \propto R_{\rm in}^{-1}$, 
we can conclude that $Q_{\rm crit}>Q$ for $M_{\ast}< 10^{3}M_{\sun}$ 
unless $R_{\rm in}$ becomes as large as $10^{3}R_{\sun}$ and 
$Q$ remains the same value.
We consider that this situation does not occur in reality 
as long as we regard $R_{\rm in}$ as the stellar surface.
Therefore, the growth of a HII region does not prevent the formation 
of such massive stars as $10^{3} M_{\sun}$.

\section{Summary and Discussion}
To estimate the mass upper limit of metal-free stars 
imposed by the formation of an HII region, 
we have found solutions for spherically symmetric and steady accretion 
flow onto a star emitting ionizing radiation.

The behavior of the flow is determined by the velocity of the flow
entering the HII region.
If the flow velocity at the edge of the HII region   
goes beyond (below) a critical value, the flow is always accelerating 
(decelerating, respectively) in the HII region.
When this value is equal to the critical value, the velocity remains
constant in the HII region.
In this critical flow, the radiation force due to ionizing photons exactly 
balances the gravity.
We applied those solutions for a compact HII region forming 
around an accreting star.  
As the HII region grows in radius, the accreting flow evolves 
from an accelerating solution to a critical one.
Soon after the critical flow is reached, accretion is halted 
by radiation force due to the ionizing radiation.
However, even stars as massive as $10^{3}M_{\sun}$ are unable to 
emit ionizing photons sufficient to halt the accretion.
The halting by gas pressure is even more difficult.
Therefore, contrary to the previous expectation, 
the formation of the HII region does not impose a stringent mass upper limit 
(at least up to $10^{3}M_{\sun}$) on metal-free stars.

We have estimated the upper limit of the mass of Pop III stars 
in relation to the formation of an HII region.
Note that other effects (e.g., stellar wind, mass outflow due to
pulsational instability, etc.) could grow in importance and
might decrease the upper mass limit of Pop III stars below the value
obtained in this paper.
We leave detailed studies on these topics for future work.

Also, in this paper, we have adopted very simplified assumptions, for example 
spherical symmetry, steady accretion, etc.
Here, we discuss other complexities and possible deviations from our picture.

We have considered only photoionization and 
electron scattering as sources of radiation force.
Here, we mention briefly the radiation force due to Ly$\alpha$ photons.
Ly$\alpha$ photons are emitted from recombination in the HII region. 
The flow inside the HII region is exerted by the radiation force due to 
Ly$\alpha$ photons.
We can extend our theory easily to include the Ly$\alpha$ pressure ,
since, in the HII region, the radiation force due to the Ly$\alpha$ photons 
is proportional to that due to ionization photons 
(equation 12 of Haehnelt 1995; 
see also Braun \& Dekel 1989; Bithell 1990).
However, according to Haehnelt(1995), the radiation force 
due to Ly$\alpha$ photons is at most of marginal importance 
relative to that due to the ionizing photons.
Thus, for the sake of simplicity,  we have not included it.
The HII region is surrounded by an HI layer, which is very optically thick 
to Ly$\alpha$ photons.
Without dust grains, which absorb Ly$\alpha$ photons and reemit
them as infrared photons, the Ly$\alpha$ photons must diffuse out 
through the HI layer. 
In the course of this, the HI layer is pushed outward by those photons.
Doroshkevich \& Kolesnik (1976) argued that 
this mechanism expels the HI layer soon after the HII region is formed 
and thereby limits the mass of stars below 10 M$_{\sun}$.
However, Harrington (1973) showed that, even without dust, the 
two-photon emission process decreases the Ly$\alpha$ photon density 
drastically, and the Ly$\alpha$ radiation force is not dynamically 
important in an HI layer surrounding an HII region.
In our case, supersonic motion in the accreting envelope also decreases 
the Ly$\alpha$ photon density.
Considering these facts, it is likely that the Ly$\alpha$ radiation force
does not play a significant role in our case.  
Therefore, we chose to neglect it here and assumed the free-fall 
outside the HII region.

We have identified the base of an HII region as the stellar surface, 
and have taken the typical value of the stellar radius $10R_{\sun}$ 
as the inner boundary radius $R_{\rm in}$ of the HII region.
However, when the accretion rate is high, a photosphere may be formed 
in the accreting flow 
(Wolfire \& Cassinelli 1986; Stahler, Palla, \& Salpeter 1986).
In this case, $R_{\rm in}$ should be taken as the radius of the photosphere,
since we assume the HII region is optically thin to continuum absorption 
except for photoionization outside $R_{\rm in}$.
In spite of higher $R_{\rm in}$, this effect generally works 
towards the smaller HII region, since $Q$ drops 
(Wolfire \& Cassinelli 1986).
Hence, our conclusion of higher upper mass limit than the former estimate 
remains the same.

The spherical symmetry is clearly an oversimplification 
if the accreting matter has large angular momentum 
and an accretion disk is formed. 
Even in this case, the spherical symmetric flow is 
a good approximation outside the centrifugal radius. 
Suppose here that inside the centrifugal radius, there is a ``cavity'' 
around the disk.
In this case, $R_{\rm in}$ should be taken as the radius of the cavity. 
If $Q$ is unattenuated inside the cavity, 
the halting of the flow outside becomes easier 
because of higher $R_{\rm in}$ and then lower $Q_{\rm crit}$ 
(see eq. [\ref{eq:qc_0}]).
The accretion through the disk might continue, however (Nakano 1989).
Those issues are still too speculative and beyond the scope of 
this paper.

Although the accelerating solution is Rayleigh-Taylor stable,
it might be unstable to 
perturbations:
if a portion of the flow becomes slightly slower than the rest, 
it becomes slower and slower relative to the average flow owing to 
the increased radiation force.
In this case, density irregularities or 
blobs could be formed in the flow, and 
our spherically symmetric solution might 
be regarded as an approximate description of the average flow.   
Further study of the stability and dynamical evolution of 
the flow will be interesting. 

\acknowledgements
We thank H.Sato for helpful comments and V.Bromm for providing the data of
Pop III ZAMS.
This work is supported in part by Research Fellowships of the Japan
Society for the Promotion of Science for Young Scientists, grant 6819.


\clearpage

\plottwo{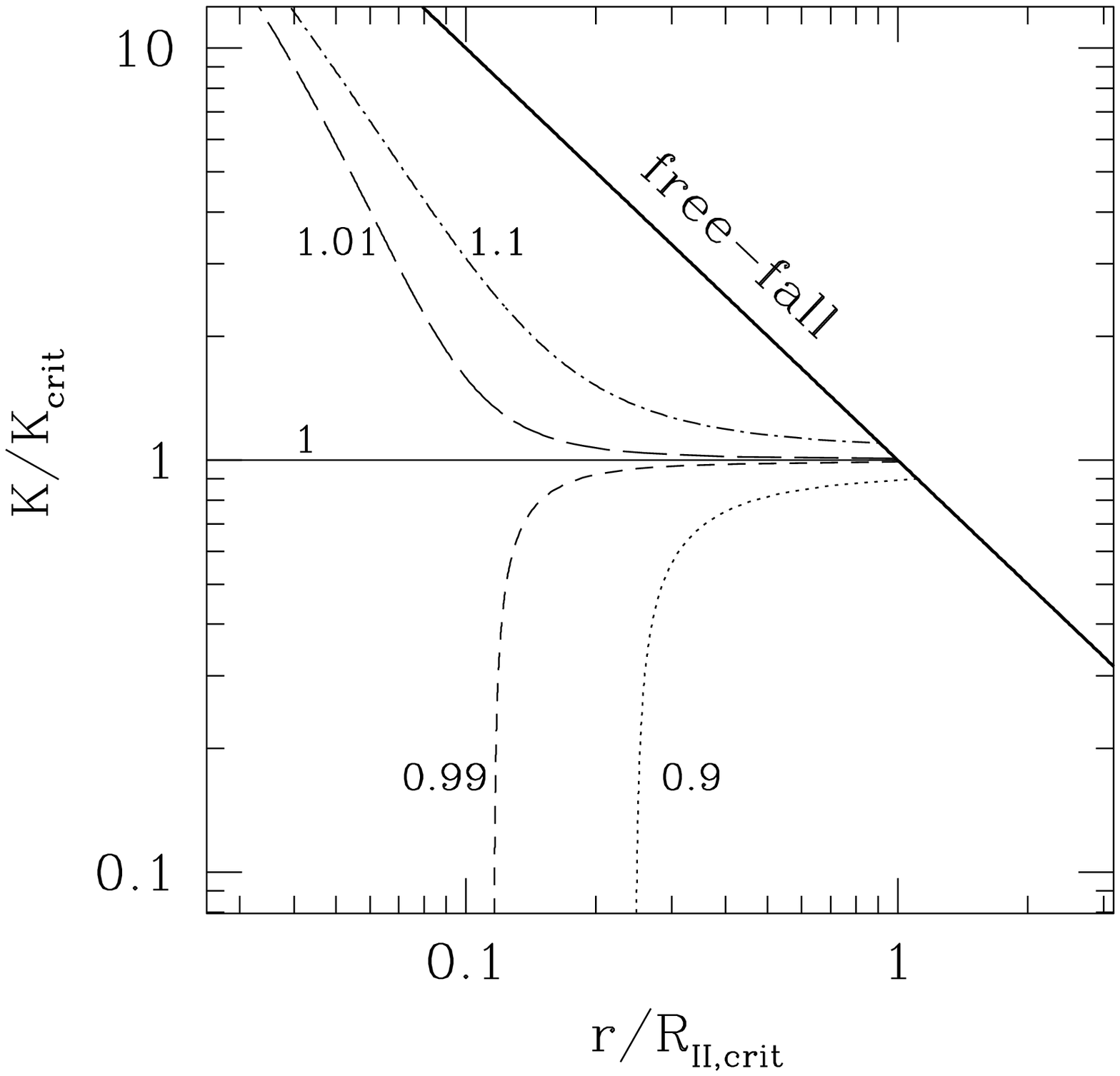}{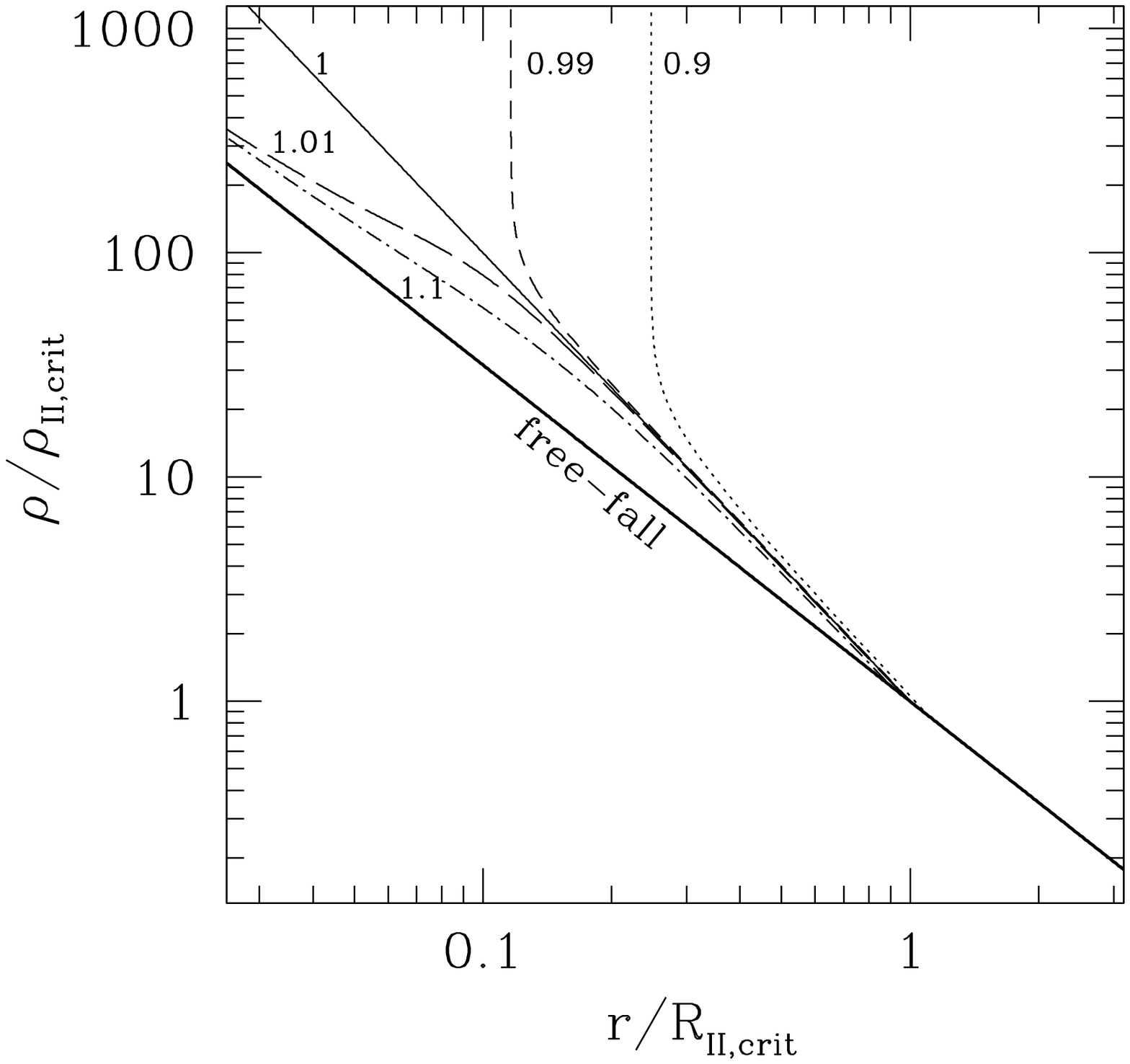}
\figcaption[f1a.eps,f1b.eps]{Solutions of accretion flows in an HII region. 
(a) the kinetic energy $K$ and (b) density $\rho$ distributions for flows
with $K_{\rm II}/K_{\rm crit}$=1.1(dash-dotted), 1.01(long-dashed),
1 (solid), 0.99(dashed), and 0.9 (dotted).
The thick solid line depicts the free-falling flow.
The density normalizer $\rho_{\rm II crit}$ is the density of 
the uniform-velocity solution (i.e., $K_{\rm II}/K_{\rm crit}=1$) 
at $r=R_{\rm II}$.
The other normalizers can be found in the text.
\label{fig1}}

\newpage
\plotone{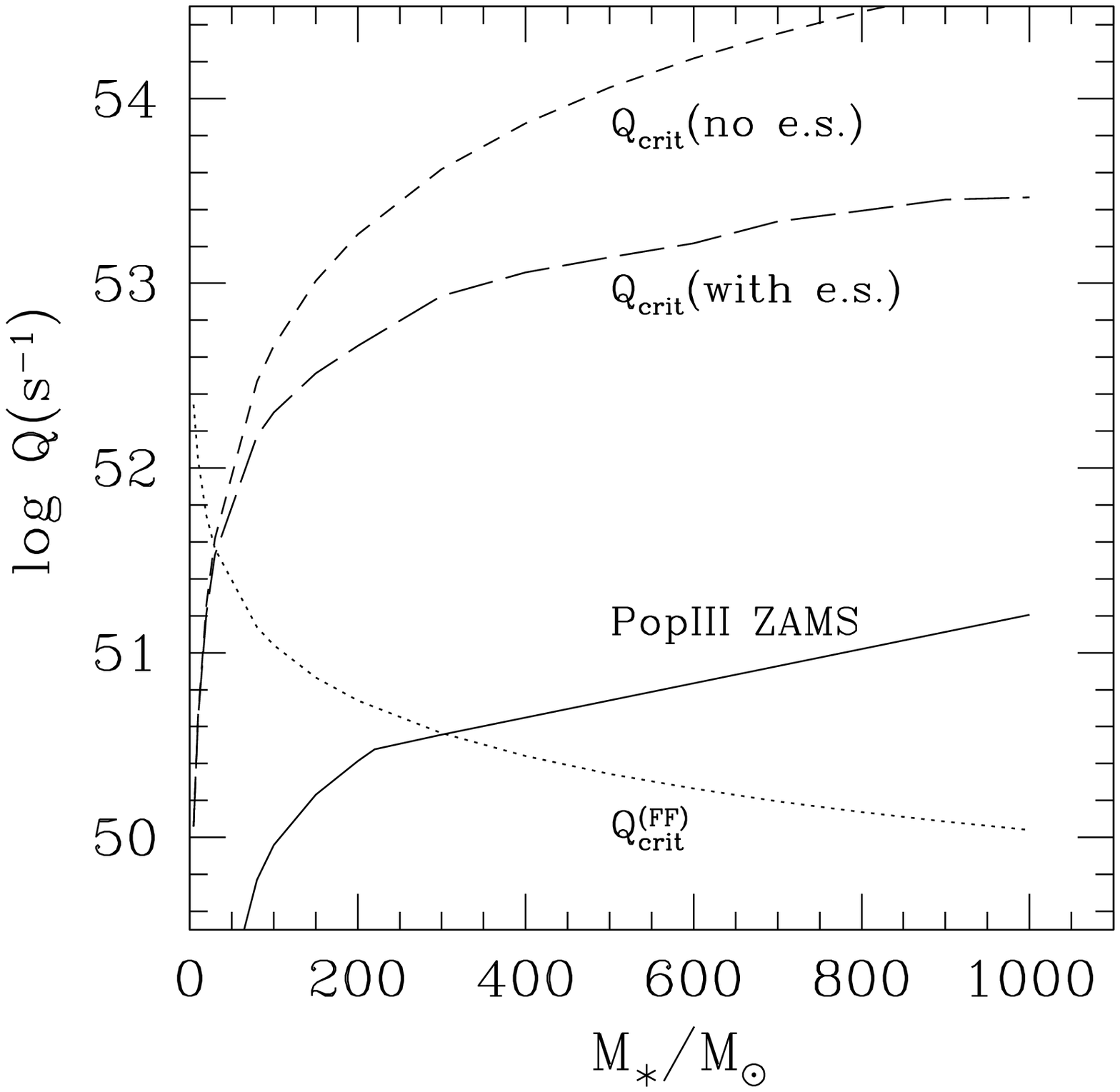}
\figcaption[f2.eps]{The ionizating photon emissivity of Pop III stars 
$Q$ vs the critical value for halting of the accretion $Q_{\rm crit}$
as a function of the stellar mass.
The solid line shows the actual ionization photon emissivity 
of ZAMS Pop III stars.
The critical values $Q_{\rm crit}$ are shown for both cases with 
(long-dashed; eq. [\ref{eq:qc_es}]) 
and without (short-dashed; eq. [\ref{eq:qc_0}]) 
the effect of electron scattering.
Also shown is the critical value $Q_{\rm crit}^{\rm (FF)}$ under the 
free-falling assumption (dotted; eq. [\ref{eq:qcrit}]).
We assume the following values of parameters:
$\dot{M}=10^{-3} M_{\sun}{\rm yr}^{-3}$ and $R_{\rm in}=10R_{\sun}$.
The values of $Q$ and $\Gamma$ are adopted from Bithell (1990) for 
$M_{\ast} < 500 M_{\ast}$ and Bromm et al.(2001) for higher masses.  
\label{fig2}}

\end{document}